\documentclass[aps,prb,twocolumn,groupedaddress]{revtex4}
\usepackage{lipsum}
\usepackage{graphicx}
\usepackage{amsfonts}
\usepackage{amsm ath}
\usepackage{amssymb}
\usepackage{graphicx}
\usepackage{epstopdf}
\usepackage{color}
\usepackage{bbm}
\usepackage{mathrsfs}

\input epsf.sty 

\newcommand{\be}{\begin{eqnarray}}
\newcommand{\ee}{\end{eqnarray}}

\begin{document}

\title{Geometry-induced topological superconductivity}
\author{Po-Hao Chou$^{1}$, Chia-Hsin Chen$^2$, Shih-Wei Liu$^{2}$, Chung-Hou Chung$^{3}$, and  Chung-Yu Mou$^{1,2,4}$
}
\affiliation{$^{1}$Physics Division, National Center for Theoretical Sciences, P.O.Box 2-131, Hsinchu, Taiwan, R.O.C.}
\affiliation{$^{2}$Center for Quantum Technology and Department of Physics, National Tsing Hua University, Hsinchu,Taiwan 300, R.O.C.}
\affiliation{$^{3}$Electrophysics Department, National Chiao-Tung University, Hsinchu,Taiwan 300, R.O.C.}
\affiliation{$^{4}$Institute of Physics, Academia Sinica, Nankang, Taiwan 115, R.O.C.}

\begin{abstract}
Intrinsic topological superconductors with $p$-wave pairing are rare in nature. Its underlying reason is due to the fact that it is usually difficult to change the relative strength between the singlet and triplet channels for the electron-electron interaction in material. Here we show that by considering superconductivity occurring on surfaces of topological insulators (TIs), the relative strength between the singlet and triplet channels can be changed by geometry and sizes of TIs. Specifically, we show that pairing of electrons at different locations on the surface of a topological insulator generally tends to favor the triplet pairing and can induce topological superconductivity by controlling the surface curvature and size of the topological insulator. We illustrate the effects in two configurations, thin film geometry and the spherical geometry with a sphere or a hemisphere, and find that topological superconductivity arises with the $p \pm ip$ pairing symmetry dominated in nanoscale size of the TI. As a consequence, vortices can spontaneously form on surfaces of topological insulators with roughness of appropriate curvature. These vortices support a Majorana zero mode inside each core and can be used as a platform to host Majorana zero modes without invoking real magnetic fields. Our theoretical discovery opens a new route to realize topological superconductivity in material.

\end{abstract}

\maketitle

\section{Introduction} 
The discovery of the topological insulators (TIs) has naturally led to consider superconductors with nontrivial topology\cite{Kanerev,Zhangrev,Zhang2009,Xue2011}. 
Due to additional particle-hole symmetry imposed by superconductivity, quasi-particles in topological superconductors can realize the Majorana fermion which is its own anti-particle with quasi-particle operators satisfying $\gamma^{\dagger} = \gamma$. In particular, the localized Majorana zero mode is considered as an important building block for constructing topological quantum computers\cite{quantumcomp}. 

To realize topological superconductivity, a number of proposals have been put forth. 
Fu and Kane proposed to realize topological superconductivity through the proximity 
of the surface states in TIs to s-wave superconductors, where in the presence of magnetic
fields, the surface Dirac cone can be used to simulate the chiral $p_x+i p_y$ pairing\cite{FuKane}.  
This idea was further developed to use other materials with spin-orbit interactions in proximity to both ferromagnetic insulators and s-wave superconductors\cite{Sarma}. While much experimental progress has been made along this approach\cite{Xu2015,William2012, Wang2012, Sun2016}, the confinement of Majorana zero modes to the interface of the TI and a superconductor limits experimental access for manipulations.  On the other hand, materials that develop intrinsic topological superconductivity generally requires the presence of the odd-parity pairing symmetry\cite{Sato2017}. In real materials, singlet and triplet pairing channels are governed by the same electron-electron interactions but projected to different channels. 
For the singlet pairing channel, the wavefuction of the Cooper pair needs to be symmetric in space; while for the triplet pairing channel, the wavefuction for the Cooper pair has to be anti-symmetric in space. The Coulomb interaction energy is reduced when the orbital wavefunction is antisymmetric in space; while the energy for the attractive interaction between electrons 
due to the electron-phonon interaction is reduced when the orbital wavefucntion is symmetric in space.  For typical superconductors, the attractive interaction between electrons dominates so that the summation of electron-phonon interaction and the Coulomb interaction yields a net attraction between electrons near the Fermi surface.   As a result, the condition for the odd-parity superconductivity to emerge is usually difficult to be satisfied. Hence natural topological superconductors are rare. The well-known old candidates are superfluid $^3$He\cite{He3} and Sr$_2$RuO$_4$\cite{SRO}. Recent intensive search has led to the discovery of a number of new possible candidates such as 
Cu$_x$Bi$_2$Se$_3$, In$_x$Sn$_{1-x}$Te, PrOs$_4$Sb$_{12}$, $\beta$-PdBi$_2$, UTe$_2$\cite{CuBiSe1,CuBiSe2, Bi2Se3pressure, Sb2Se3pressure, InSnTe, CuPbSeBiSe, SrBi2Se3, Tl5Te3, PrOs4Sb12, betaPdBi2, UTe2}, and etc.  Although there are experimental evidences of unconventional superconductivity in these materials, due to possible alternative scenario to account for experimental observations\cite{Lee2012, SRO1}, unambiguous signature of topological superconductivity has not been firmly established. 

In this paper, by considering the surface superconductivity that was recently found on TIs\cite{Sb2Te3}, we would like to explore an alternative way to realize the topological superconductivity. 
By exploiting the competition of the Coulomb interaction and the electron-electron attraction due to phonons on surfaces, we show that the relative strength of the Coulomb interaction and the electron-electron attraction due to phonons can be tuned by curvatures and sizes of TI. Specifically, we show that through pairing of electrons at different locations on surfaces of a TI,  surfaces of TIs can be turned into topological superconductors by controlling surface curvature and sizes of TIs. We illustrate the mechanism by first considering competition of pairing of electrons on the same surface and paring between top and down surfaces  (interlayer) in the thin film geometry, and we then extend the illustration to the same competition on a curve surface by using the spherical geometry with a sphere and a hemisphere. In the thin film geometry, we find that the inter-layer pairing with $p \pm ip$ pairing symmetry emerges when the thickness decreases to the nanoscale. In the spherical geometry, we show that  $p \pm ip$ pairing dominates and vortices can spontaneously form without invoking real magnetic fields. These vortices generally supports a Majorana zero mode inside each vortex core. Our results thus indicate that TIs with appropriate geometry (such as surface roughness) can be used as a platform to host Majorana zero modes on their surfaces.

\section{Theoretical model} 
We start by considering electrons in a general manifold $M$ occupied by a topological insulator. The low energy physics is characterized by interacting Dirac Hamiltonian
\begin{eqnarray} \label{Eq1}
H =&\int_{M} d\vec{r} \ \psi^{\dag} ({\vec{r}})
\begin{pmatrix} (m_{\vec{r}}-\mu) I & \lambda_{so}\vec{\sigma}\cdot \vec{p} \notag \\
           \lambda_{so}\vec{\sigma}\cdot \vec{p}  & (-m_{\vec{r}}-\mu) I
\end{pmatrix}\psi ({\vec{r}}) \\
&+\frac{1}{2}\int_{M} d\vec{r} d\vec{r'} V (\vec{r},\vec{r}') \psi^{\dag} (\vec{r}) \psi (\vec{r}) \psi^{\dag} (\vec{r}') \psi (\vec{r}'). \nonumber  
\label{4DiracH}
\end{eqnarray}
Here $\psi^{\dag} (\vec{r}) =(C_{\vec{r},A\uparrow}^{\dag},C_{\vec{r},A\downarrow}^{\dag},C_{\vec{r},B\uparrow}^{\dag},C_{\vec{r},B\downarrow}^{\dag})$ with $C_{\vec{r},\tau\sigma}^{\dag}$ being the creation operator of conduction electrons with orbital $\tau=A,B$\cite{Zhang2010} and spin $\sigma=\uparrow,\downarrow$ at position $\vec{r}$. $\mu$ is the chemical potential. $m_{\vec{r}}=m_0+m_2\vec{p}^2$
is the effective Dirac mass with $\vec{p}$ being the momentum operator, $m_0>0$, and $m_2<0$.  The interaction $V(\vec{r},\vec{r}')$ includes the phonon-mediated attractive interaction $V^{ph} (\vec{r},\vec{r}')$ and the repulsive Coulomb interaction $V^{c} (\vec{r},\vec{r}')$. We shall assume that the superconductivity occurs in topological protected surface states so that $H$ will be projected onto the curved surface of $M$ and become an effective surface Hamiltonian $H_{eff}$. The energy dispersion of $H_{eff}$ includes Dirac cones with particle-hole symmetry. Here the chemical potential $\mu$ and the superconducting pairing energy cutoff will be set to lie in lower Dirac cone.
\section{Topological $p$-wave pairing induced in thin film geometry} 
\begin{figure}[hbtp] 
\includegraphics[height=2.0in, width=3.0in]{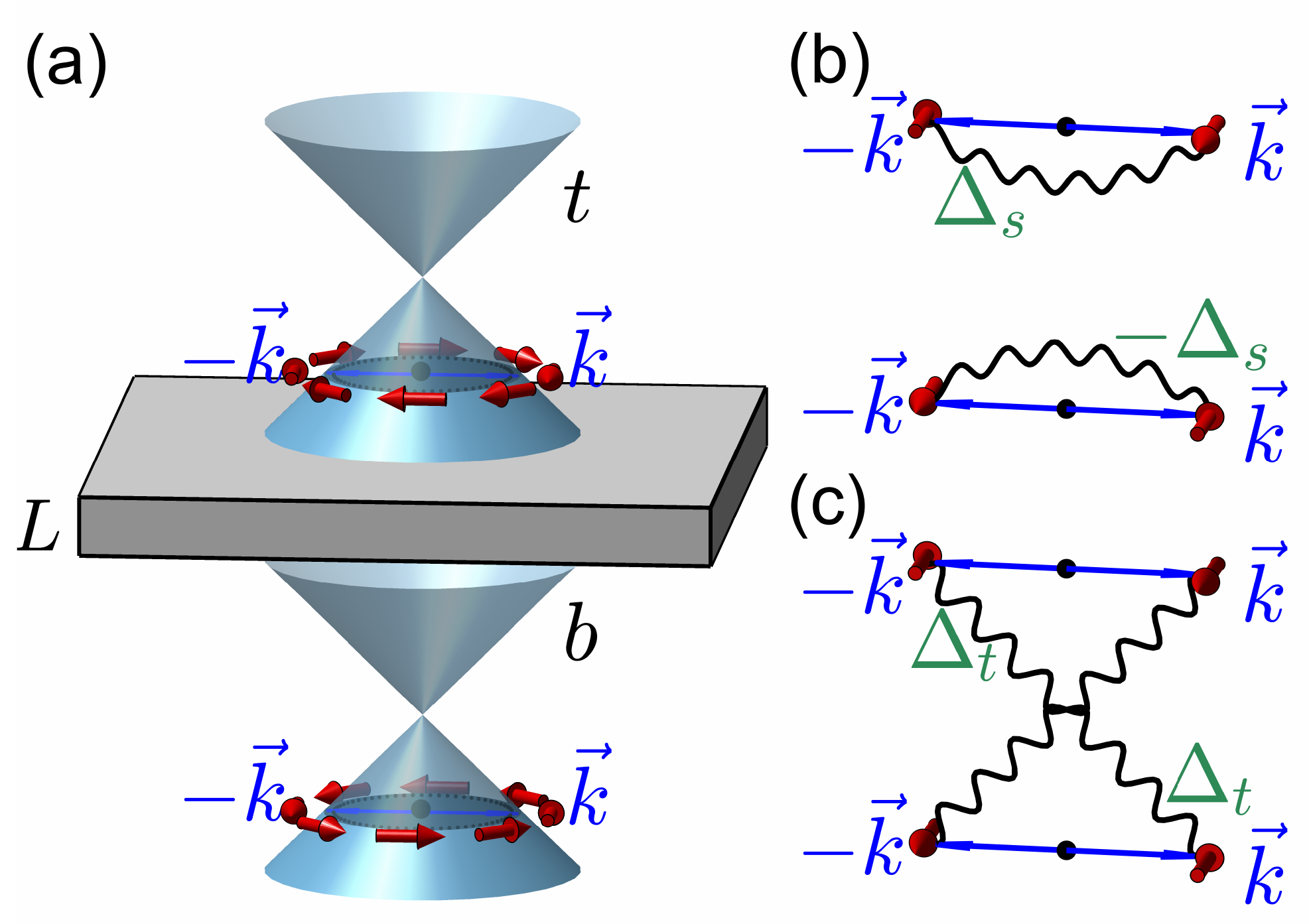}
\centering
\caption{(a) Schematic plot of Dirac cone states on top (t) and bottom (b) surfaces in a thin film of thickness L. (b) Pairing of electrons in the same surface, $\Delta_s$ (top surface) and $-\Delta_s$ (bottom surfaces), is singlet. (c) Pairing of electrons between top and bottom surfaces, $\Delta_t$, is triplet.}\label{Fig1}
\end{figure}
We first analyze superconductivity due to the competition between $V^{ph}$ and $V^c$ in the thin film geometry.
Let the thickness of the thin film be $L$. As shown in Fig.~\ref{Fig1}(a), electronic states at top and down surfaces are described by the Dirac cone states when $L$ is large. The possible pairing of electrons are intra-surface and inter-surface pairings as indicated in ~\ref{Fig1}(b) and (c). Clearly, the combined intra-surface pairing of top ($t$) and bottom ($b$) surfaces is singlet-pairing, represented by $\Delta_s$; while the  inter-surface pairing is triplet-pairing, represented by $\Delta_t$.   The Hamiltonian for describing the surface superconductivity is the specialization of Eq.(\ref{Eq1}) to the thin film geometry and can be written as
\begin{eqnarray}
&& H= \sum_{\mathbf{k} z}  \Psi^{\dagger}_{{\mathbf k}z} (H_{\mathbf{k}z}-\varepsilon_F) \Psi_{{\mathbf k}z}+ \nonumber \\
 && \frac{1}{2\Omega}  \sum_{\mathbf{k} \mathbf{k}' \mathbf{q}, zz'} V_{\mathbf{q},zz'}( \Psi^{\dagger}_{{\mathbf k}-{\mathbf q}z}\Psi_{{\mathbf k}' z})( \Psi^{\dagger}_{{\mathbf k}'+{\mathbf q}z'}\Psi_{{\mathbf k} z'}),
\end{eqnarray}
where $\varepsilon_F$ is the Fermi energy,  $\Omega$ is the surface area, A and B are indices for atomic orbits, $z$ is the index for quintuple layers along z-axis,  ${\mathbf k}=(k_x,k_y)$ is the in-plane momentum along the surface, and $\Psi^{\dagger}_{{\mathbf k}z} = ( C^{\dagger}_{{\mathbf k}z,A \uparrow}, C^{\dagger}_{{\mathbf k}z,A \downarrow},C^{\dagger}_{{\mathbf k}z,B \uparrow}, C^{\dagger}_{{\mathbf k}z,B \downarrow})$ is the creation operator for surface electrons. The Hamiltonian matrix $H_{\mathbf{k}z}$ is given by the following form\cite{Zhang2010}
\begin{eqnarray} \label{Dirac2}
&& H_{\mathbf{k}z}= (m_0+m_1 p^2_z + m_2 k^2 ) \tau_z+ \nonumber \\
&&  \left[ \lambda_{so}^z p_z \sigma _z + \lambda_{so} \left(k_x \sigma_x+k_y\sigma_y  \right) \right]\tau_x ,
\end{eqnarray}
where $\tau_{\alpha}$ ($\alpha=x,y,z$) are Pauli matrices in the pseudo space formed by atomic orbits A and B and parameters are given as follows: $m_0 =-175$ (meV), $m_1=200$ (meV$ \cdot \mbox{nm}^{2}$), $m_2=500$ (meV$\cdot \mbox{nm}^2$),  $\lambda_{so} = 330$ (meV$\cdot \mbox{nm}$) and $\lambda_{so}^z = 330$ (meV$\cdot \mbox{nm}$). The interaction
$V_{\mathbf{q},zz'}$  is a summation of bare Coulomb interaction and phonon-mediated interaction,$V_{\mathbf{q},zz'}=V^c_{\mathbf{q},zz'}+V^{ph}_{\mathbf{q},zz'}$, with
\begin{eqnarray}
&&V^c_{\mathbf{q},zz'}=\frac{4\pi e^2}{\kappa} \frac{e^{-q|z-z'|}}{2q}, \nonumber \\
&&V^{ph}_{\mathbf{q},zz'}=-\frac{(Z\hbar q \overline{V^c_q})^{2}}{Ma^2} \sum_{\varepsilon}  \frac{\phi_{\varepsilon} (z) \phi_{\varepsilon} (z')}{(v_p q)^2+\varepsilon^2}.
\end{eqnarray}
Here $\overline{V^c_q}$ is the average of  $V^c_{\mathbf{q},zz'}$, $Z=3.53e$ is the renormalized  ion charge, $M=124.7u$ is the average ion mass, $a=0.4$ nm is the lattice constant along the surface,  $v_p=2$ km/s is the speed of sound, $\varepsilon$ is the phonon dispersion calculated by 1D phonon model with open boundary condition along z-axis, and $\phi_{\varepsilon} (z)$ is the corresponding wave function of phonon.

There are four surface energy eigenstates for each Fourier mode ${\mathbf k}$, which are related to the real space basis by a unitary transformation $U$
\begin{equation}
C^{\dagger}_{{\mathbf k},v \alpha} = \sum_{z \tau \sigma} U^{\tau \sigma *}_{{\mathbf k},v \alpha} (z) C^{\dagger}_{{\mathbf k} z,\tau \sigma} .
\end{equation}
Here $v=\pm$ specifies the spin rotation direction of the surface state, i.e. the chirality of the surface state, $\alpha=p,n$  specifies sign of the eigen-energy(positive or negative),  $\tau=A,B$ specifies atomic orbits, and $\sigma$ is the index for spin. We shall assume that only the negative energy surface states lie on the Fermi level, hence only the $\alpha=n$ channel is kept and we shall drop the index for $\alpha$. The screened Coulomb interaction and phonon-mediated interaction, $V^c_{\mathbf{q},zz'}$ and $V^{ph}_{\mathbf{q},zz'}$,  can be obtained from the z-axis charge screening as done in Ref[\onlinecite{screened}]. The resulting BCS Hamiltonian is then given by
\begin{eqnarray}\label{Eq2}
&&H_{eff}=\sum_{\mathbf{k}}\sum_{v=\pm}(\varepsilon_k-\mu)C^{\dag}_{\mathbf{k},v}C_{\mathbf{k},v} + \frac{1}{2 \Omega} \times\notag \\
&&\sum_{\begin{subarray}{c}\mathbf{k}\mathbf{k}' \\v_1, v_2,v_3, v_4 = +,-\end{subarray}}
\Gamma^{v_1v_2v_3v_4}_{\mathbf{k}\mathbf{k}'}C^{\dag}_{\mathbf{k},v_1}C^{\dag}_{-\mathbf{k},v_2}C_{-\mathbf{k}',v_3}C_{\mathbf{k}',v_4}.\quad
\end{eqnarray}
Here $\Omega$ is the surface area of the film, $\varepsilon_k = - \sqrt{\lambda_{so}^2k^2+\Delta^{2}_{g,k}}$ with $\Delta_{g,k}$ being the hybridized gap of top and bottom surface states, $v=\pm$ is the index for chirality, and $\Gamma^{v_1v_2v_3v_4}_{\mathbf{k}\mathbf{k}'}$ is the projected effective interaction between quasi-particles of different chiralities.
\begin{figure}[t]
\includegraphics[height=2.5in,width=3.5in]{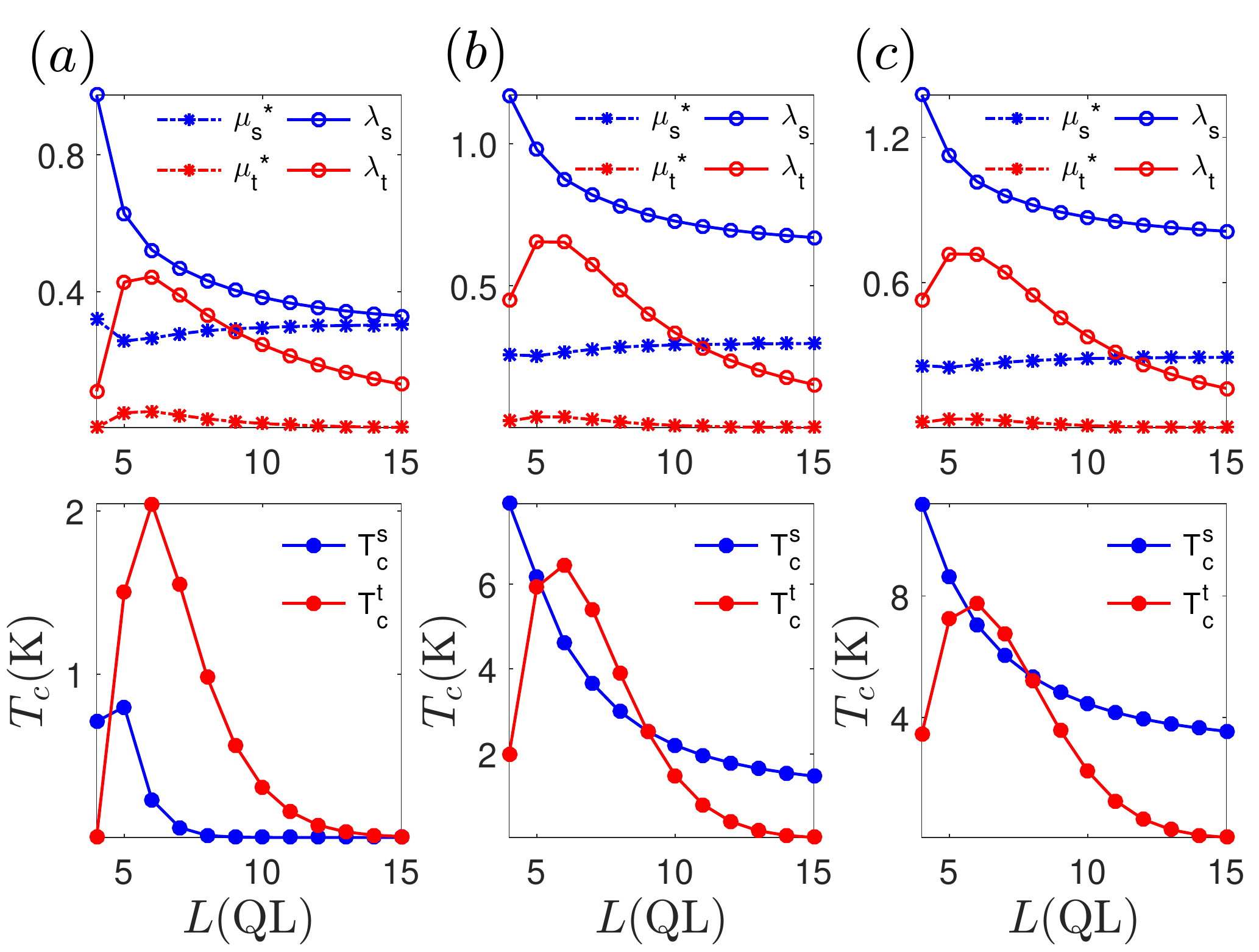}
\caption{Thickness dependence of electron-phonon coupling $\lambda_{\alpha}$, strength of Coulomb interaction $\mu^*_{\alpha}$, and $T_c$ for three different Fermi energies\cite{parameter1}: (a) $0.5\varepsilon_F$ (b) $0.84 \varepsilon_F$ (c) $\varepsilon_F$, where $\varepsilon_F$ is the bulk Fermi energy of $\mbox{Sb}_2\mbox{Te}_3$\cite{Sb2Te3}. Here $L$ is number of quintuple layers(QL, 1 QL $\sim$ 1nm).  Subscripts $s$ and $t$ denote intra-surface and inter-surface.}\label{Fig2}
\end{figure}
The pairing amplitudes, $\Delta_s$ and $\Delta_t$, are determined by the effective potential by $g_{s,\mathbf{k}\mathbf{k}'}=2(\Gamma^{++++}_{\mathbf{k}\mathbf{k}'}-\Gamma^{++--}_{\mathbf{k}\mathbf{k}'})$ and $g_{t,\mathbf{k}\mathbf{k}'}=2(\Gamma^{+--+}_{\mathbf{k}\mathbf{k}'}+\Gamma^{+-+-}_{\mathbf{k}\mathbf{k}'})$. In the mean-field approximation, the pairing term in $H_{eff}$ becomes 
\begin{eqnarray}\label{Eq3}
&&H_{\Delta} =\sum_\mathbf{k}  \left\{ -i \Delta_s (C^{\dag}_{\mathbf{k}\uparrow}C^{\dag}_{-\mathbf{k}\downarrow}- C^{\dag}_{\mathbf{k}\downarrow}C^{\dag}_{-\mathbf{k}\uparrow})+\Delta_t (k_x-ik_y ) \right. \nonumber \\
&& \left. C^{\dag}_{\mathbf{k}\uparrow}C^{\dag}_{-\mathbf{k}\uparrow} -\Delta_t (k_x+ik_y ) C^{\dag}_{\mathbf{k}\downarrow}C^{\dag}_{-\mathbf{k}\downarrow}+h.c. \right\}.
\end{eqnarray}
As a result, the energy of quasi-particles is given by  $E_k=\sqrt{\xi^{2}_{k}+\Delta^{2}_{s}+\Delta^{2}_{t}}$ with $\xi_{k}=\varepsilon_{k}-\mu$ and the gap equations are given by $\Delta_{\alpha}=-\sum_{\mathbf{k}'}g_{\alpha,\mathbf{k}\mathbf{k}'}
\frac{\Delta_{\alpha}}{2E_{k'}}\tanh{\frac{\beta E_{k'}}{2}},$
where $\alpha$= $t$ or $s$ and $\beta=1/k_BT$. 
Clearly, the gap equation cannot be simultaneously satisfied for both $\Delta_t$ and $\Delta_s$. Hence $\Delta_t$ and $\Delta_s$ cannot coexist for a given thickness $L$. Since interactions between quasi-particles crucially depend on $L$, their competition may lead to the change of pairing symmetry as the thickness changes. An estimation  of the transition temperature can be found based on the empirical McMillan formula\cite{McMillan}
\begin{equation}
T_c =\frac{\Theta_D}{1.45} \exp \left[ - \frac{1.04(1+\lambda)}{\lambda-\mu^* (1+0.62\lambda)} \right].
\end{equation}
\begin{figure}[hbtp] 
\includegraphics[height=1.8in, width=2.5in]{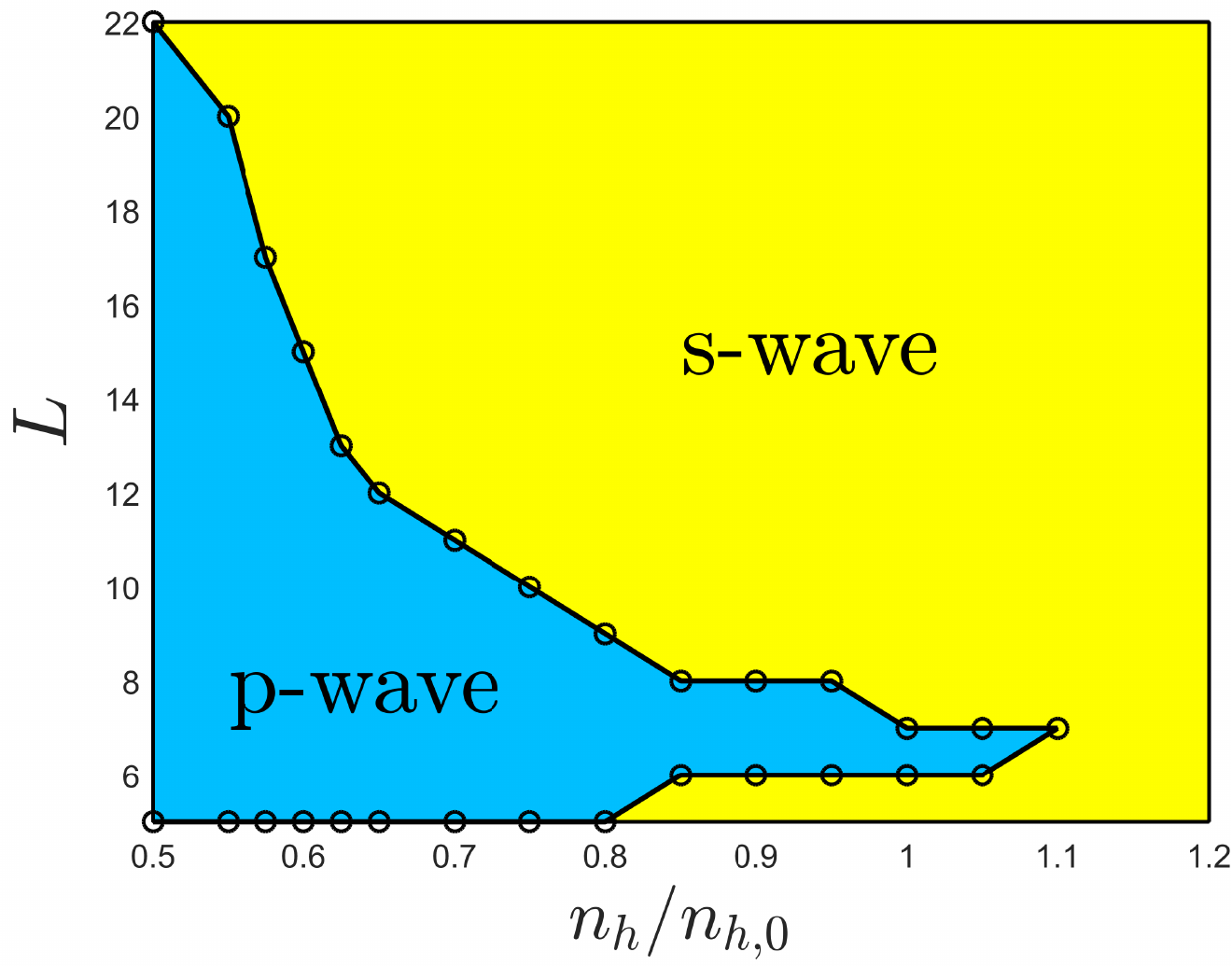}
\centering
\centering
\caption{Phase diagram of superconducting orders for thin film geometry with the Fermi energy being in the lower surface Dirac cone. 
Here $L$ is the thickness of the thin film in the unit of number of quintuple layers, $n_h$ is the surface hole density, and $n_{h,0}$ is the bulk hole density of $\mbox{Sb}_2\mbox{Te}_3$. }
\label{phase_diagram}
\end{figure}[hbtp]
Here $\Theta_D$ is the Debye temperature, $\lambda$ is the electron-phonon coupling constant, and $\mu^*$ is the Coulomb pseudo-potential. $\lambda$ and $\mu^*$ are obtained by setting $k,k' \sim k_F$ and performing the average over the angle of $k'$, $\phi_{{\mathbf k}'}$ ,for the effective potential due tothe electron-phonon coupling 
$ g^{ph}_{\alpha,{\mathbf k} {\mathbf k}'}$and the Coulomb interaction $ g^{c}_{\alpha,{\mathbf k} {\mathbf k}'}$ respectively
\begin{eqnarray}
&&\lambda_{\alpha} = \int_{0}^{2\pi} d  \phi_{{\mathbf k}'} g^{ph}_{\alpha,{\mathbf k}_F {\mathbf k}_F'}, \nonumber \\
&&\mu_{\alpha} = \int_{0}^{2\pi} d  \phi_{{\mathbf k}'} g^{c}_{\alpha,{\mathbf k}_F {\mathbf k}_F'},\nonumber \\ && \mu^*_{\alpha}=\frac{\mu_{\alpha}}{1+\mu_{\alpha}\ln{(W/\Theta_D)}},
 \end{eqnarray}
where $\alpha = s$ or $t$, and $ W \sim \varepsilon_F$ is the Coulomb interaction cut off . Fig.~\ref{Fig2}(a) shows $\lambda_{\alpha}$ and $\mu^*_{\alpha}$ versus $L$ with values $\lambda_s$ at large $L$ being consistent with experiments\cite{lambda}. In Fig.~\ref{Fig2}(b), we find that the triplet pairing wins over for $L \sim 5$-$10$ nm for larger $\varepsilon_F$\cite{phase}, indicating that pairing between parallel surfaces may enhance the triplet pairing and turn a TI into a surface topological superconductor in thin film geometry. Finally, the resulting phase diagram of superconducting orders in thin films with the Fermi energy being in the lower surface Dirac cone is given in Fig.~\ref{phase_diagram}.
\section{Effective magnetic field induced by curvature}  The results of thin film geometry imply that pairing of electrons at different locations on a curved surface may also enhance the triplet pairing.  Hence we consider superconductivity on a general surface.  It is well known
that the Dirac equation on a curved space acquires the spin connection as a gauge field\cite{Nakahara}.
In particular, it is found that a spin-1/2 Fermion on a sphere in the local frame will see an effective magnetic
monopole $\pm 1/2$ for the local spin direction being $\mp$ respectively\cite{monopole}.  We have generalized this result to a general surface $S$ and proved that the effective local magnetic field $\vec{B}_{eff}$ is proportional to the Gauss curvature $K$ as \cite{general_monopole}
\begin{equation}
\vec{B}_{eff} = \pm \frac{1}{2} K \hat{n},
\end{equation}
\begin{figure}[hbtp] 
\includegraphics[height=1.2in, width=3.2in]{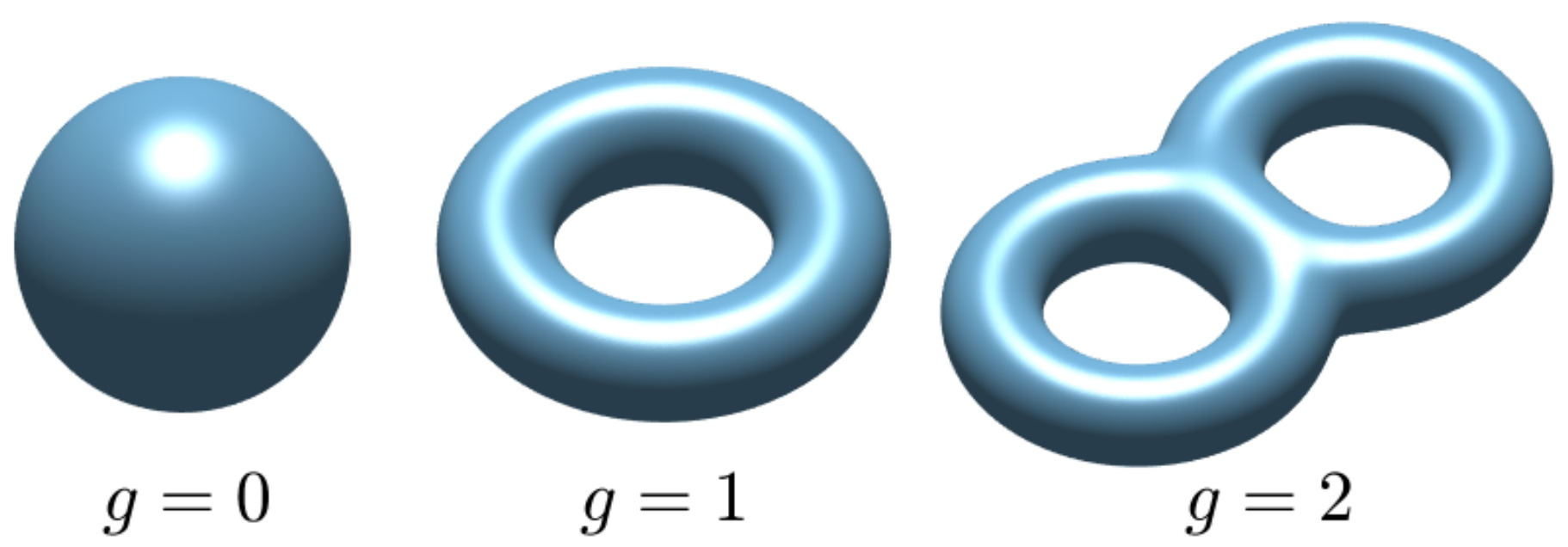}
\centering
\centering
\caption{The total magnetic flux due to the spin-connection gauge field that passes through a surface $S$  is a topological invariant, which depends on the genus $g$ of the surface as $\oint_S \vec{B}_{eff} \cdot d \vec{a} = \pm 2 \pi(1-g)$, where $\pm$ are the local spin directions (down and up) of electrons.} \label{Fig3}
\end{figure}
where $ \hat{n}$ is the unit vector normal to the surface and $\pm$ are the local spin directions of electrons. As a result, according to the Gauss-Bonnet theorem, the total magnetic flux through a surface $S$, $\oint_S \vec{B}_{eff} \cdot d \vec{a}$, is a topological invariant that depends only the genus $g$ of $S$ as illustrated in Fig.~\ref{Fig3}. 
For a general surface with genus $g$, considering possible formation of vortices, the Poincar\'{e}-Hopf theorem\cite{poincare} further implies that the minimum number of vortices on the surface is $2 (1-g)$. The formation of vortices is determined by  $\vec{B}_{eff}$ and energetics with the configuration of the lowest free energy being realized. 
\section{Curvature induced superconducting vortex states}  
To consider effects of effective magnetic fields on surface superconductivity, we shall consider the simplest surface with $g=0$. In particularly, we shall consider the surface on a sphere  or a semi-sphere with different curvature.  As we shall see, the results allow one to further estimate the formation of vortices on general surfaces.  In the presence of effective magnetic fields induced by curvature,  since electrons with different spin directions see opposite monopole charges, we expect that only Cooper pairs of the triplet pairing, $++$ and $--$, are affected by non-vanishing $B_{eff}$, while the singlet pairing would not be affected by $B_{eff}$.  The effective Hamiltonian is given by
\begin{eqnarray}\label{Eq4}
&& H_{eff}=\int d\Omega \Psi^\dag_{\Omega}(\mathcal{H}_{surf}-\mu)\Psi_{\Omega}+ \nonumber  \\
&& \frac{1}{2}\int\int d\Omega d\Omega'V(\Omega,\Omega') (\Psi^\dag_{\Omega}\Psi_{\Omega})(\Psi^\dag_{\Omega'}\Psi_{\Omega'}).
\end{eqnarray}
Here $\Psi^\dag_{\Omega}\equiv (C^\dag_{\Omega,+} , C^\dag_{\Omega,-})$ is the creation operator for the local spinor at the solid angle $\Omega$. $\mathcal{H}_{surf}$ is the projective Hamiltonian describing the free local surface states
\begin{eqnarray}
\mathcal{H}_{surf} = \frac{\lambda_{so}}{R}\Bigg( \begin{array}{cccc}
                  0 & D_- \\
                  D_+ & 0
             \end{array}\Bigg),
\end{eqnarray}
where $R$ is the radius of sphere and $D_{\pm} =e^{\pm i\phi}(\pm \partial_{\theta}+\frac{i}{\sin\theta}\partial_{\phi} \mp \frac{1}{2}\tan\frac{\theta}{2})$\cite{local}. $V(\Omega,\Omega')$ is the effective interaction that includes the screened Coulomb interaction adapted to the sphere, $V^c (\Omega,\Omega')=\frac{V_C}{2R \sin \frac{\Delta\Omega}{2}} \exp (-\alpha_c |\sin (\frac{\Delta\Omega}{2})|) $ and the attractive phonon-mediated interaction  $V^{ph}(\Omega,\Omega')=-V_{ph} \exp (-\alpha_{ph}(\Delta\Omega)^2)$ with $\Delta\Omega=\Omega-\Omega'$. Here the interactions are modeled by a Gaussian form\cite{SC_sphere} and the Yukawa potential form with $\alpha_{ph}=\frac{R}{\xi_{ph}}$  and $\alpha_c=\frac{2R}{\xi_C}$ ($\xi_{ph}$ and $\xi_C$  are the corresponding decay length and screening length). Finally, as the attractive phonon interaction is cutoff in energy, we assume that $\mathcal{H}_{surf}$ is cutoff in energy near the chemical potential with the cutoff $\Lambda$.

The mean-field pairing amplitude is given by $\Delta^{s,s'}_{\Omega,\Omega'} =V(\Omega,\Omega')\langle C_{\Omega,s}C_{\Omega',s'} \rangle$ so that the pairing Hamiltonian is $H_{\Delta} = \frac{1}{2}\int\int d\Omega d\Omega' \sum\limits_{(s,s')}\Delta^{s,s'}_{\Omega,\Omega'}C^\dag_{\Omega,s}C^\dag_{\Omega',s'}+h.c.$.
To solve the self-consistent gap equation on the sphere, we need to go to the eigen-basis of ${H}_{surf}$ with the eigen-energies being $E_{\pm}= \pm(\lambda_{so}/R)(j+1/2)$ with $j=0,1,2,\cdots$. The corresponding annihilation operators are denoted by $C_{jm,E_{\pm}}$, which are related to the local spinor $C_{\Omega \pm}$ by 
$C_{jm,E_{\pm}}=\int d\Omega (Y^{m*}_{-1/2,j}C_{\Omega +} \pm Y^{m*}_{1/2,j}C_{\Omega -})$. Here $Y^{m}_{q,j}$ are the monopole harmonics with $q$ being the monopole charge and $(j,m)$ being the angular momentum quantum number. By keeping the $E_{-}$ operator and dropping the subindex $E_{\pm}$, we obtain
\begin{eqnarray}\label{eigen1}
\begin{array}{l}
C_{jm}^\dag  = \frac{1}{{\sqrt 2 }}\int {d\Omega (Y_{ - 1/2,j}^m(\Omega )C_{\Omega,+ }^\dag  - Y_{1/2,j}^m(\Omega )C_{\Omega,- }^\dag )}, \\
C_{\Omega,+ }^\dag  = \frac{1}{{\sqrt 2 }}\sum\limits_{jm} {C_{jm}^\dag Y_{ - 1/2,j}^{m * }(\Omega )}, \\
C_{\Omega,- }^\dag  =  - \frac{1}{{\sqrt 2 }}\sum\limits_{jm} {C_{jm}^\dag Y_{1/2,j}^{m * }(\Omega )}.
\end{array}
\end{eqnarray}
The interaction $V(\Omega ,\Omega ')$ can be re-expressed in this basis. First,
we set $\gamma  = \Omega - \Omega^\prime$ and apply the addition theorem of spherical harmonics to write $V(\Omega,\Omega^{\prime})$ as
\begin{eqnarray}
V(\Omega ,\Omega ')
= \sum\limits_l {\frac{{4\pi V_l}}{{2l + 1}}\sum\limits_{\bar m} {{{( - 1)}^{\bar m}}Y_l^{ - \bar m}(\Omega )Y_l^{\bar m}(\Omega ')} },
\end{eqnarray}
where
\begin{eqnarray}
{V_l} = \frac{{2l + 1}}{2}\int_{ - 1}^1 {V(\gamma){P_l}(\cos{\gamma})d\cos{\gamma}},
\end{eqnarray}
$P_l(x)$ is the $l_{th}$ Legendre polynomial, and $Y_l^{\bar m}$ is the spherical harmonics.
By using Eq.(\ref{eigen1}), the interaction term in Eq.(\ref{Eq4}) can be expressed as
\begin{eqnarray}
&&\int\int{d\Omega d\Omega 'V(\Omega ,\Omega ')(\Psi _\Omega ^\dag {{ \Psi }_\Omega })
( \Psi _{\Omega '}^\dag {{ \Psi }_{\Omega '}})} \nonumber \\
&=&\frac{1}{2}\int\int{d\Omega d\Omega 'V(\Omega ,\Omega ')(C_{\Omega,+ }^\dag {C_{\Omega,+ }}{\rm{ + }}C_{\Omega,- }^\dag {C_{\Omega,- }})} \times \nonumber \\
&&(C_{\Omega ',+ }^\dag {C_{\Omega ',+ }}{\rm{ + }}C_{\Omega ' ,- }^\dag {C_{\Omega ' ,- }}) \nonumber \\
=&&\frac{1}{2}\sum_{\begin{subarray}{c}j_1j_2j_3j_4 \\ m_1m_2m_3m_4\end{subarray}}
V_{j_1j_2j_3j_4}^{m_1m_2m_3m_4}C^{\dag}_{j_1m_1}C^{\dag}_{j_2m_2}C_{j_3m_3}C_{j_4m_4}, \nonumber \\
\end{eqnarray}
where
\begin{eqnarray}
&&V_{j_1j_2j_3j_4}^{m_1m_2m_3m_4}= \nonumber \\
&& \int\int d\Omega d\Omega' \sum_{l}\frac{\pi V_{l}}{(2l+1)}\sum_{\bar m}(-1)^{\bar m}Y_l^{-\bar m}(\Omega )Y_l^{\bar m}(\Omega ') \nonumber\\
&&\times \sum_{q,q'=\pm 1/2} Y_{q,j_1}^{m_1*}(\Omega)Y_{q,j_4}^{m_4}(\Omega)Y_{q',j_2}^{m_2*}(\Omega')Y_{q',j_3}^{m_3}(\Omega').
\end{eqnarray}
The integrals involved between three harmonics are given by\cite{sup4}
\begin{eqnarray}\label{3j}
&&\int {d\Omega } {Y_l^{\bar m}(\Omega)Y_{q,j}^{m*}(\Omega)Y_{q,j'}^{m'}(\Omega)}  \nonumber \\
= &&{( - 1)^{l + j + j' - q -  m}}{\left[ {\frac{{(2l + 1)(2j + 1)(2j' + 1)}}{4 \pi}} \right]^{1/2}} \nonumber  \\
&&\times \left( {\begin{array}{*{20}{c}}
l&j&{j'}\\
{\bar m}&{ - m}&{m'}
\end{array}} \right)\left( {\begin{array}{*{20}{c}}
l&j&{j'}\\
0&{ - q}&q
\end{array}} \right),
\end{eqnarray} where $\left( {\begin{array}{*{20}{c}}
l&j&{j'}\\
{\bar m}&{ - m}&{m'}
\end{array}} \right)$ is the $3-j$ symbol. Furthermore, to ensure that the anti-commuting properties of fermion operator are preserved in the interaction, we define
\begin{eqnarray}
&&\Gamma_{j_1j_2j_3j_4}^{m_1m_2m_3m_4}=\frac{1}{4}(V_{j_1j_2j_3j_4}^{m_1m_2m_3m_4}-V_{j_2j_1j_3j_4}^{m_2m_1m_3m_4}
 \nonumber \\
&&-V_{j_1j_2j_4j_3}^{m_1m_2m_4m_3}+V_{j_2j_1j_4j_3}^{m_2m_1m_4m_3}).
\end{eqnarray}
Eq.({\ref{Eq4}) is then turned into the following form
\begin{eqnarray}\label{Eq5}
&&H^{surf}_{eff}=\sum_{jm}\left[-\lambda_{so}/R(j+1/2)-\mu \right]C_{jm}^{\dag}C_{jm} \notag \\
&&+\frac{1}{2} \sum_{\begin{subarray}{c}j_1j_2j_3j_4 \\ m_1m_2m_3m_4\end{subarray}}
\Gamma_{j_1j_2j_3j_4}^{m_1m_2m_3m_4}C^{\dag}_{j_1m_1}C^{\dag}_{j_2m_2}C_{j_3m_3}C_{j_4m_4},\quad \nonumber \\
\end{eqnarray}
while the corresponding mean-field equations are given by
\begin{eqnarray}
\Delta^{mm'}_{jj'}=\sum_{\begin{subarray}{c}j_3j_4 \\ m_3m_4\end{subarray}}\Gamma_{jj'j_3j_4}^{mm'm_3m_4}\langle C_{j_3m_3}C_{j_4m_4} \rangle,
\end{eqnarray}
where  $\Gamma_{j_1j_2j_3j_4}^{m_1m_2m_3m_4}$ is the interaction $V(\Omega,\Omega')$ expressed in the $(j,m)$ basis. 

To find the superconducting state, we first solve $\Delta^{mm'}_{jj'}$. The  solution  is then converted to the solid-angle basis, $\Delta^{s,s'}_{\Omega,\Omega'}$. The gap function generally depends on solid angles $\Omega$ and $\Omega'$ of two electrons in a Cooper pair and can be decomposed into the superposition of products of  the amplitude at the center of mass point  $\bar{\Omega}= (\Omega + \Omega')/2$ and the amplitude for the relative motion that presents different pairing symmetries as $
\Delta^{s,s'}_{\Omega,\Omega'} = \Delta_S (\bar{\Omega}) S(\Omega, \Omega') + \Delta_P  (\bar{\Omega}) P(\Omega,\Omega')+ \cdots,$
where $S(\Omega, \Omega'), P(\Omega, \Omega'), \cdots$ denote pairing symmetries of  s-wave, p-wave, $\cdots$ with the corresponding amplitudes $\Delta_s (\bar{\Omega}),  \Delta_p  (\bar{\Omega}), \cdots$ at the center of mass point.
The s-wave, $S(\Omega, \Omega')$, is a constant. The local pairing symmetry, $p_x \pm ip_y$, on a sphere is taken to be in the form and their higher harmonics in solid angles\cite{SC_sphere}: $(P+iP) (\Omega,\Omega')= \alpha \beta' -\beta \alpha' $ and 
$(P-iP) (\Omega,\Omega')= \alpha \beta'^{*} -\beta \alpha'^{*}$, where ($\alpha$, $\beta$) denotes the local spinor\cite{local,local1}. Numerical values of the pairing amplitude can be obtained by first taking the limit $\Omega -\Omega' \rightarrow 0$ to find amplitudes of relative motion using the different form of pairing symmetries \cite{SC_sphere_2}. The coefficients that go with different pairing symmetries are the amplitudes at the center of mass point when $\Omega =\Omega'$.
\begin{figure*}[thbp]
  \includegraphics[width=6.0in,height=3.1in]{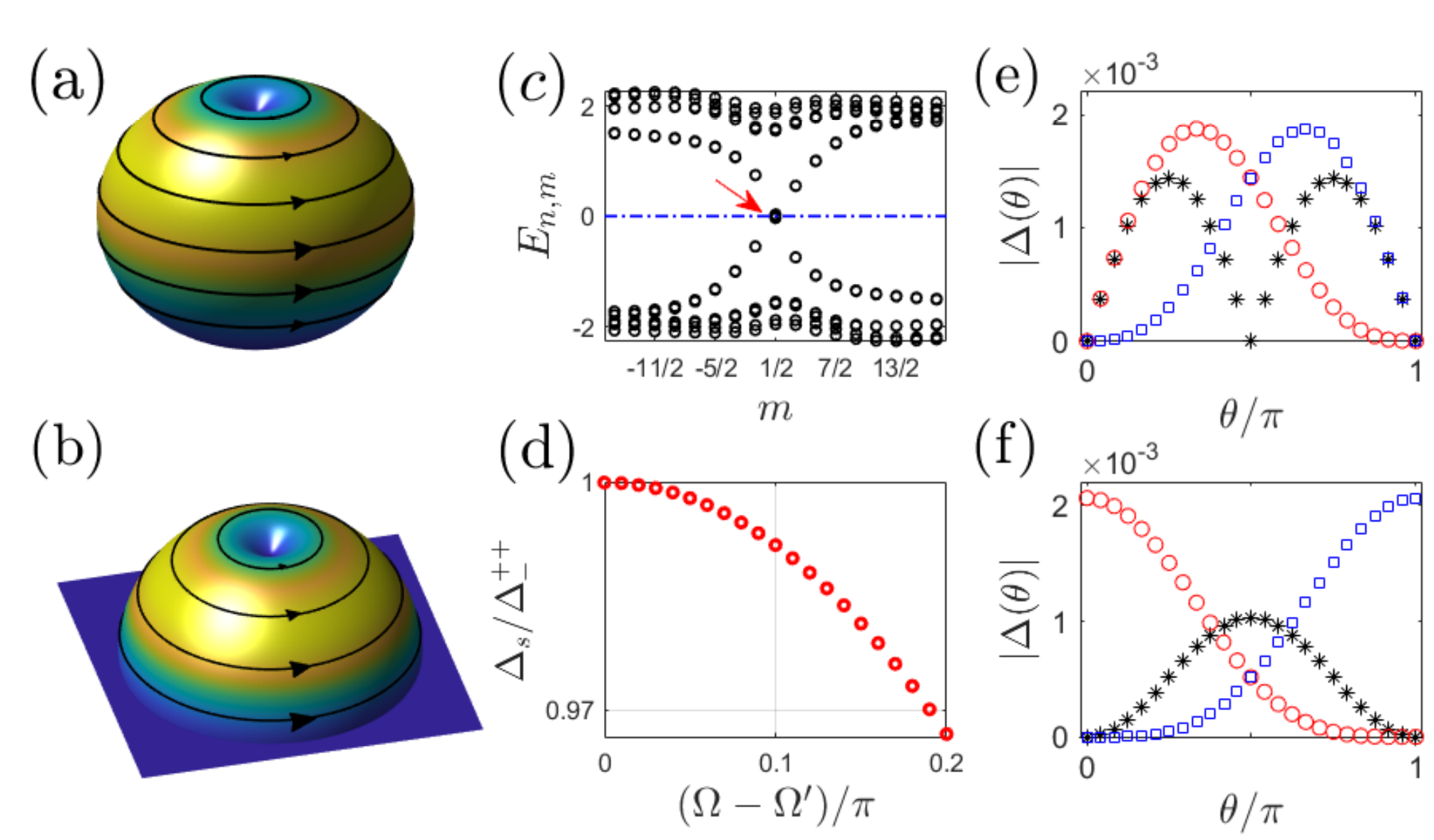}
 \caption{Self-consistent vortex solution on the surface of a spherical TI and a semi-spherical TI. Here $\lambda_{so}=35$meV$\cdot$nm, $\xi_C=4$nm, $\xi_{ph}=50$nm, $R=50$ nm, $V_{ph}=7.76$meV, $V_C=50$ meV$\cdot$nm, $\mu=-9.5$meV, the cutoff $\Lambda=2.1$meV, and the electron density is $2.7\times 10^{-3}$nm$^{-2}$\cite{parameter2}. (a) The superconducting amplitude, $\Delta^{++}$ (meV) (pairing symmetry $p_x$+$ip_y$), on a sphere for the vortex solution with vortex charge $Q=1$ in the  vortex phase. Here the color-scale decreases from 0.001meV(yellow)  to 0(blue).  Contours and arrows mark directions of supercurrents.  (b) The superconducting amplitude, $\Delta^{++}$ (meV), for a vortex solution on a semi-sphere. (c) The quasi-particle energy spectrum (in the unit of meV) inside the vortex core exhibits the Majorana zero mode (indicated by the arrow, tolerance is 0.02meV). Here $m$ and $n$ are z-component of angular momentum of the quasi-particle. (d) Relative strength of the singlet pairing to the triplet pairing ($\Delta^{++}_-$) versus $\Omega- \Omega'$ in the Meissner phase. Here $V_C=70$ meV and $T=1.24$meV.(e) and (f): Pairing components in $\Delta^{++}$ and $\Delta^{--}$ for vortex phase (e) and vortex* phase (f). Here $\theta$ is the azimuthal angle relative to the north pole, red circles represent $\Delta^{++}_{+}$, blue squares represent $\Delta^{--}_{-}$, and black dots represent  $\Delta^{++}_{-}$ (=$\Delta^{--}_{+}$).} \label{Fig4}
\end{figure*}


In Fig.~\ref{Fig4}, we show numerical solutions of generic superconducting states found on spheres and semi-spheres.  It is found that
 vortices of two different charges ($Q=1$ and $Q=2$) are spontaneously formed in different regimes\cite{vortex}. Fig.~\ref{Fig4}(a) shows $\Delta^{++}$ with the pairing symmetry $p_x+ip_y$ in color-scale for the vortex solution with one vortex of charge $Q=1$ at the north pole and the other vortex of charge $Q=1$ at the south pole. Phase with such vortices is termed as the vortex phase. A similar solution with minor component exhibiting the vortex solution of charge $Q=2$ is also found.  Phase with such vortices is termed as the vortex* phase. This is consistent with the Poincar\'{e}-Hopf theorem which states that the total winding number (= $Q$) of a vector field on a sphere should be two. 
In Fig.~\ref{Fig4}(b), we show that the spontaneously-formed vortex solution survives on a semi-sphere, which simulates a bump on the surface. This solution implies that vortices can also form on bumps with appropriate curvature on surfaces of TIs.
Fig.~\ref{Fig4}(c) shows that each vortex in the vortex phase support a Majorana zero mode in the core. For the vortex* phase, since the vortex solution only exists in the minor component ($\Delta^{++}_{-}$, see Fig.~\ref{Fig4}(f)), the vortex core does not host Majorana zero mode\cite{Sato}. Fig.~\ref{Fig4}(d) shows the relative strength of singlet pairing to triplet pairing versus $\Omega- \Omega'$ in the Meissner phase in which there is no vortex. Here only $\Delta^{++}$ and $\Delta^{--}$  ($=\Delta^{++}$) are non-vanishing triplet components in the Meissner phase. Clearly, due to the momentum-locking of Dirac fermions at different solid angles, the triplet pairing has higher chance of occurrence and dominates. This results in the topological superconductivity.  The triplet component for the vortex phase can be generally expressed as $\Delta^{++} = \Delta^{++}_{-} (p_x-ip_y)+ \Delta^{++}_{+} (p_x+ip_y)$ and  $\Delta^{--} = \Delta^{--}_{-} (p_x-ip_y)+ \Delta^{--}_{+} (p_x+ip_y)$. The components for different pairing symmetries are shown in Fig.~\ref{Fig4}(d) and (e). These components $\Delta^{ss}_{s'}$ can be expressed in terms of monopole harmonics. 

As the local magnetic field is determined by $1/R^2$, the interplay between the Meissner phase and vortex phases on a sphere is generally controlled by the radius and the strength of Coulomb interaction. Fig.~\ref{Fig5} shows typical phase diagrams we find for superconducting states on a sphere. It is seen that the vortex phase generally exists at low temperature on surfaces with radius of curvature being around the nanoscale (smaller than 50 nm). 
\begin{figure}[hbtp] 
\includegraphics[height=2.0 in, width=3.5in]{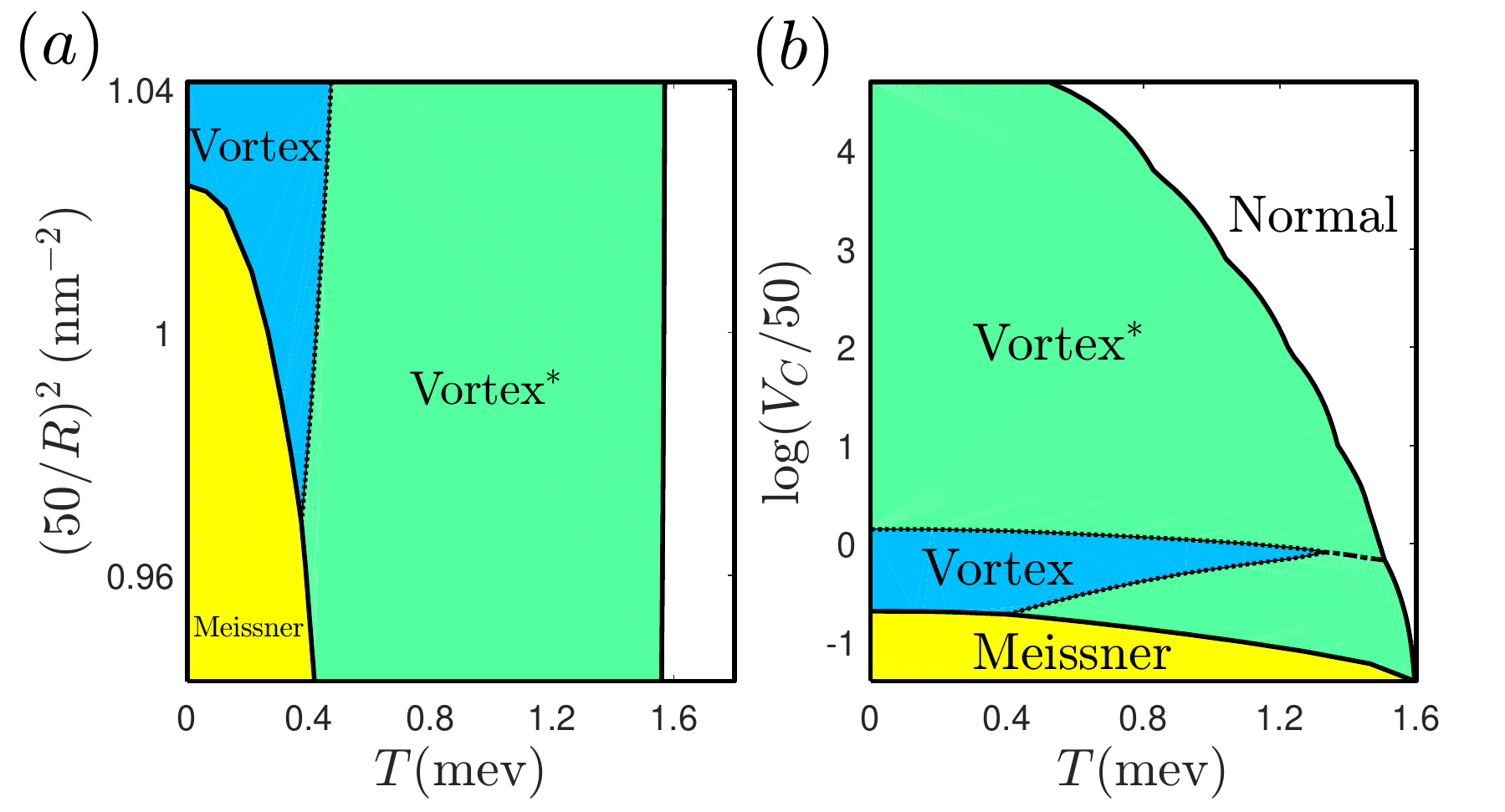}
\centering
\caption{(a) The phase diagram of superconducting states on a sphere: $1/R^2$ versus temperature $T$. Here $1/R^2$ act similar as the local magnetic field. (b) The phase diagram of superconducting states on a sphere:  $V_C$ versus $T$. Here $\xi_C=4$nm, $\xi_{ph}=50$nm, $V_{ph}=7.76$meV, $\mu=-9.5$meV, and the cutoff $\Lambda=2.1$meV. Note that vortex phases diminish as  $R \rightarrow \infty$, which can be accessed in (b) with $V_C \rightarrow 0$.}\label{Fig5}
\end{figure}

\section{Discussion and conclusion} 
Although the emergence of vortices with Majorana zero mode is mainly illustrated on a perfect sphere, our results are not restricted to a perfect sphere. This is illustrated in Fig.~\ref{Fig4}(b) by a semi-sphere, which simulates a bump on a plane. The solution also contains a Majorana zero mode inside the vortex core. It implies that vortices can also spontaneously form on bumps with appropriate curvature on surfaces of TIs. On general surfaces, the vortex state is competing with the Meissner phase in energy. At the critical curvature when the vortex state with a single vortex wins over the Meissner phase, the free energy for the vortex state is equal to that of the Meissner phase. Similar to lower critical field$H_{c1}$ for conventional superconductors, this determines the critical curvature $K_c$  for the formation of a single vortex as 
\begin{equation}
K_c =\pm  \frac{8\pi \epsilon}{\Phi_0},
\end{equation}
where $\epsilon$ is the free energy per unit length of the vortex and $\Phi_0$ is the flux quantum. Hence there is a critic radius of curvature for forming a vortex. Below the critical radius of curvature, the Meissner phase wins.  The numerical values for $K_c$ can be estimated by using the phase digram shown in Fig. ~5(a), where at different temperatures, the critical curvature for a vortex to form is determined by the boundary between the Meissner phase and the vortex phase. Our results can be thus applied to an ellipsoid or bumps on surface of topological insulator in which vortices with Majorana modes could emerge at places where the local curvature ($~1/R^2$) exceeds the threshold as marked in Fig. ~5(a).  

The connection of the thin-film geometry to the spherical geometry is in the fact that they all belong to the $g=0$ surfaces. In this case, the Poincar\'{e}-Hopf theorem implies that the minimum number of vortices is two, but generally, number of vortices in a vortex state with charge 1 can be 2, 4,6, $\dots$. The extra vortices that deviate from 2 contain vortex pairs of $\pm 1$ charges so that the Poincar\'{e}-Hopf theorem is satisfied.  For the thin-film (slab-like) geometry, it also has the genus number, $g = 0$.  Depending on curvatures of the corners, if they exceed the threshold curvature, vortices can form. The formation of vortices will appear at corners. However, minimum number of vortices that can appear is still two. In this case, one still has to control the curvature of the corner to enable the formation of vortices.

In conclusion, we have shown that pairing of electrons on different locations of  a TI can generally induce topological superconductivity through the control of surface curvature and size of the TI. These are illustrated both in the thin film geometry and in the spherical geometry. The generic phase diagram of topological superconductivity on a sphere is constructed. For general surfaces, our results imply that vortices can spontaneously formed on bumps with appropriate curvature on surfaces of TIs.  These vortices generally host a Majorana zero mode inside each core, which is shown to be robust against the presence of the disorder potential. We expect that the experimental detection can be performed on surfaces of  Sb$_2$Te$_3$\cite{Sb2Te3} by using STM and one looks for surface roughness of size around 50 nm or less at temperatures below 9K. The surface of a TI can be thus used as a platform to host Majorana zero modes without invoking real magnetic fields.  

\begin{acknowledgments}
This work was supported by the Ministry of
Science and Technology (MoST), Taiwan. We also acknowledge support
from the Center for Quantum Technology within the
framework of the Higher Education
Sprout Project by the Ministry of Education (MOE) in Taiwan.
\end{acknowledgments}


\begin{thebibliography}{99}
\bibitem{phase} $\varepsilon_F$ is determined by surface electron/hole density. Hence the general superconducting phase diagram for thin films depends on surface electron/hole densities and thickness.
\bibitem{local} This choice is based on the choice of the local spinor as $
|+\rangle \equiv \left( \begin{array}{c}
                  \alpha \\ \beta \end{array}  \right) = \left( \begin{array}{c}
                  \cos(\theta/2) \\ \sin(\theta/2)e^{i\phi} \end{array}  \right),
|-\rangle \equiv \left( \begin{array}{c}
                 - \alpha^* \\ \beta^* \end{array}  \right)  = \left( \begin{array}{c}
                  -\sin(\theta/2)e^{-i\phi} \\ \cos(\theta/2) \end{array}\right)$, see Ref. [\onlinecite{monopole}].
\bibitem{local1} The form of the paring symmetry $p_x \pm ip_y$ given here is with respect to the local frame.  There is also a spin-connection gauge field induced by the locally-defined $p_x \pm ip_y$ wave, which must be included in computing the super-current. See Ref. [\onlinecite{SC_sphere_2}].
\bibitem{vortex} For a general surface, the formation of a single vortex is determined by the condition: the Gibbs free energy for the vortex state is equal to the Gibbs free energy for the Meissner phase.  Formally, this determines the threshold curvature $K_c$ for the formation of a single vortex as $K_c =\pm  8\pi \epsilon/\Phi_0$, where $\epsilon$ is the free energy per unit length of the vortex and $\Phi_0$ is the flux quantum. Practically, $\epsilon$ is not known and one can use the phase digram shown in Fig.~ 5 to estimate the threshold curvature for formation of a single vortex on general surfaces.
\end{thebibliography}
\end{document}